\documentclass[graybox]{svmult}

\usepackage{mathptmx}       
\usepackage{helvet}         
\usepackage{courier}        
\usepackage{type1cm}        
\usepackage{makeidx}         
\usepackage{graphicx}        
\usepackage{multicol}        
\usepackage[bottom]{footmisc}

\makeindex             

\begin{document}

\title*{New Links between Pulsation and Stellar History}
\author{Nancy Remage Evans}
\institute{Nancy Remage Evans \at SAO, 60 Garden St, MS 4 Cambridge MA 02138, USA \email{nevans@cfa.harvard.edu}}
%
%
\maketitle

\abstract*{New instrumentation is providing new insights into  
intermediate mass pulsating Cepheids, particularly 
about their formation and history.  Three
approaches are discussed, using space (Hubble and Chandra) and
ground-based studies (radial velocities).
 First, we are conducting a survey of Cepheids with the Hubble Space Telescope Wide
Field Camera 3 (WFC3) to identify possible resolved companions (for
example Eta Aql)  and thus provide constraints on star formation.  
Followup  X-ray observations (Chandra and XMM-Newton) can
confirm whether possible low mass companions are young enough to be physical
companions of Cepheids.
In a related study of intermediate mass stars, Chandra X-ray
observations of late B stars in  Tr 16 have been used to
determine the fraction which have X-ray active low mass companions. 
Finally, the Tennessee State Automatic Spectroscopic Telescope AST and the 
Moscow University group have obtained velocities of a number of
Cepheids.  As an example, the orbit of V350 Sgr has been redetermined,
providing a new level of accuracy to the orbital velocity amplitude,
which is needed for mass determination.   }


\abstract{New instrumentation is providing new insights into  
intermediate mass pulsating Cepheids, particularly 
about their formation and history.  Three
approaches are discussed, using space (Hubble and Chandra) and
ground-based studies (radial velocities).
 First, we are conducting a survey of Cepheids with the Hubble Space Telescope Wide
Field Camera 3 (WFC3) to identify possible resolved companions (for
example Eta Aql)  and thus provide constraints on star formation.  
Followup  X-ray observations (Chandra and XMM-Newton) can
confirm whether possible low mass companions are young enough to be physical
companions of Cepheids.
In a related study of intermediate mass stars, Chandra X-ray
observations of late B stars in  Tr 16 have been used to
determine the fraction which have X-ray active low mass companions. 
Finally, the Tennessee State Automatic Spectroscopic Telescope AST and the 
Moscow University group have obtained velocities of a number of
Cepheids.  As an example, the orbit of V350 Sgr has been redetermined,
providing a new level of accuracy to the orbital velocity amplitude,
which is needed for mass determination. }

\vskip .1truein

\section{Introduction} 

This contribution focuses on two aspects of binary Cepheids:  
information they provide about star formation and about stellar 
evolution (masses). 

\begin{figure}
\centering
\includegraphics[scale=.20]{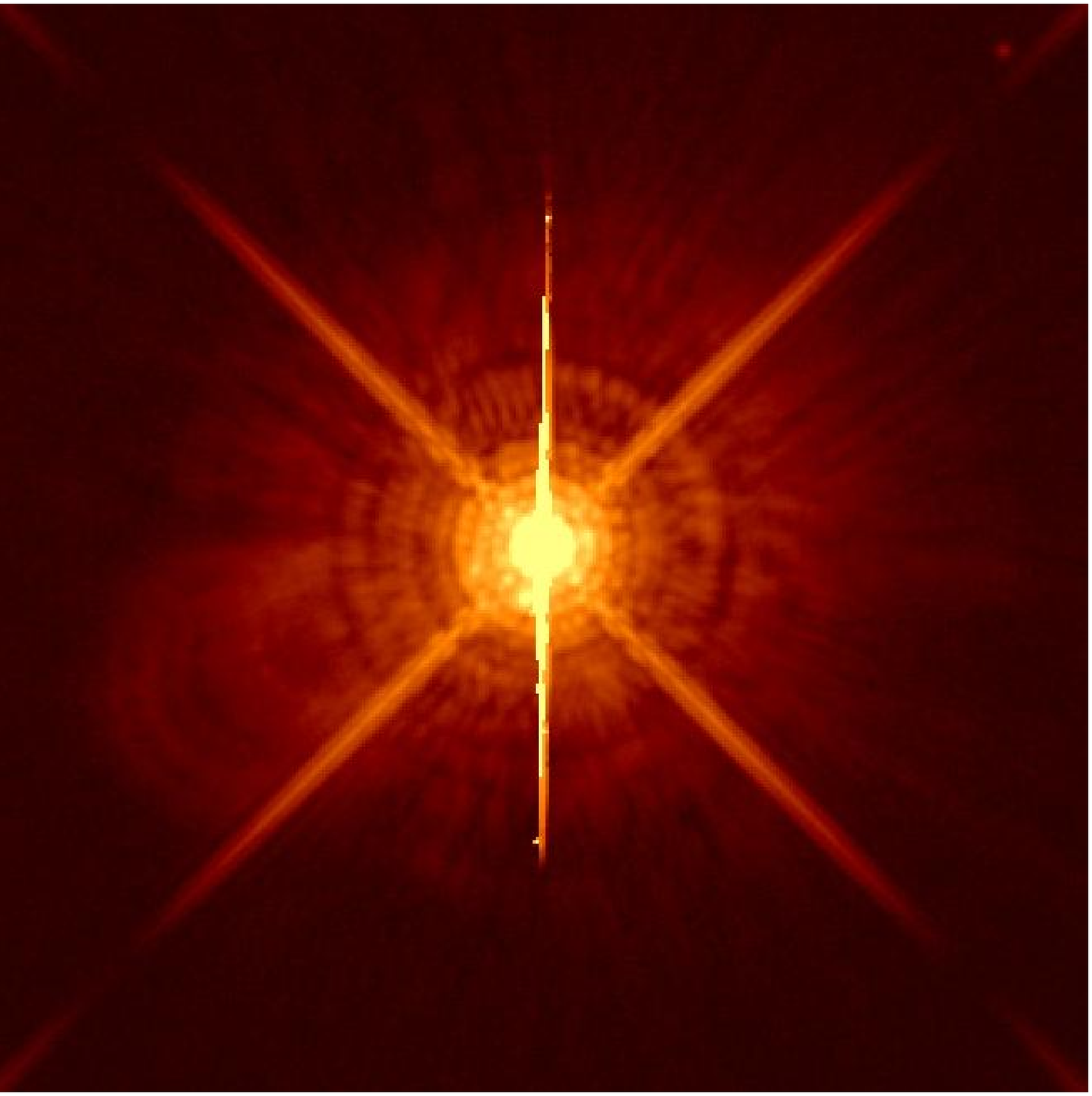}
\hspace{0.5cm}
\includegraphics[scale=.25]{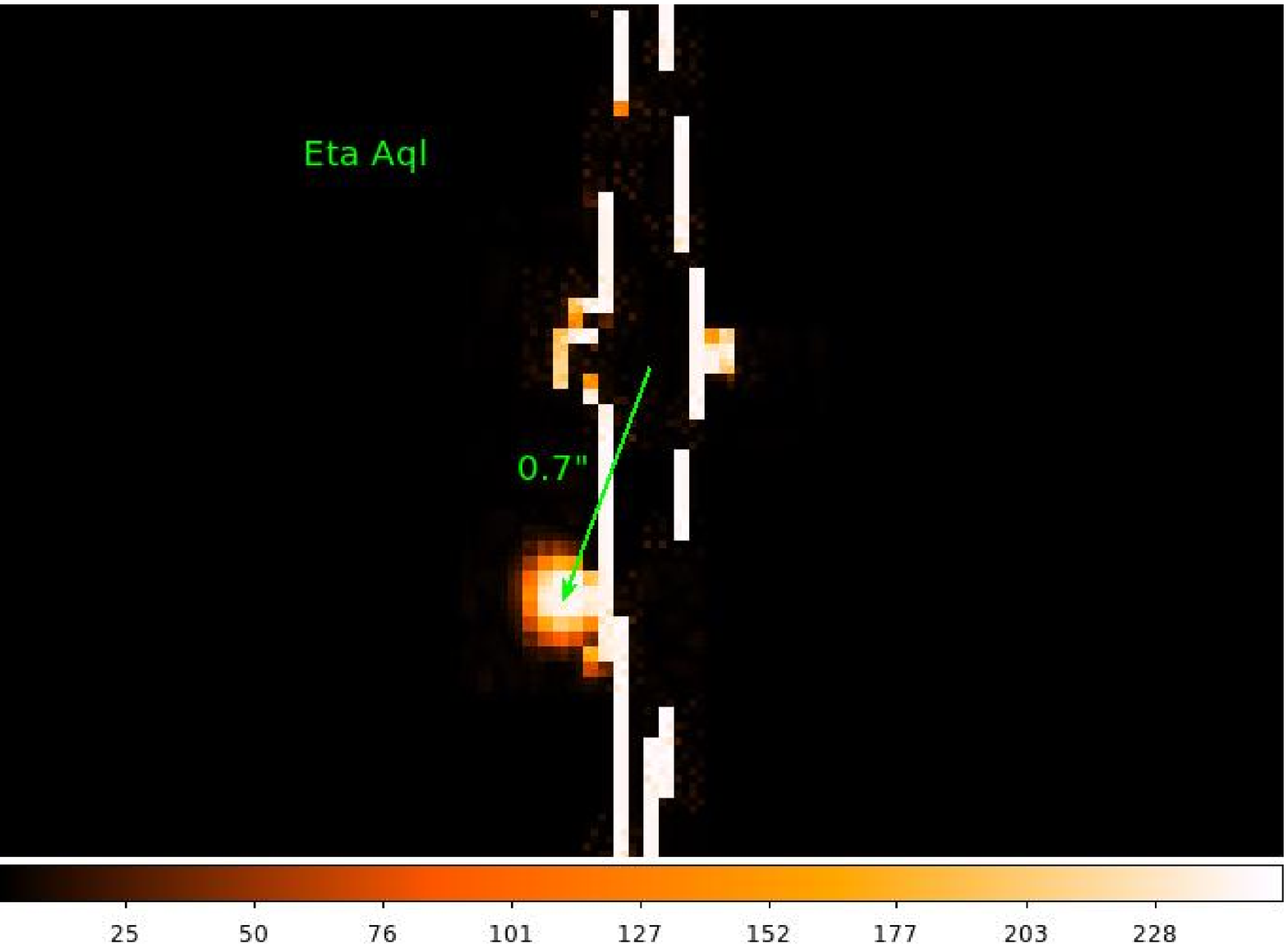}


\caption{Left: The center of the HST WFC3 V image of $\eta$ Aql.  
The image has a log scale and is approximately 10$''$ wide. Right: The difference 
image with the Cepheid T Mon subtracted from the $\eta$ Aql image.  The companion 
is clearly visible 0.7$''$ from the Cepheid.  Only the vertical column bleeding 
remains uncorrected.   }
\label{fig:1}       
\end{figure}

\section{Star Formation}
\label{sec:2}

The formation of binary/multiple systems is a very effective way to 
manipulate angular momentum as a 
cloud collapses and stars are formed.  Thus the distribution of binary 
parameters provides ``fingerprints'' of 
star formation, although, of course, some parameters may subsequently be 
altered by interactions with other stars.  For solar-mass stars, binary and 
multiple properties have been well characterized, particularly in the seminal 
study of Duquennoy and Mayor \cite{dm} which combined CORAVEL radial velocities 
with visual binaries and common proper motion pairs.  It was recently updated by
Raghavan et al. \cite{raghaven} to include recent high resolution techniques.  For more 
massive stars binary properties are less well determined, since they are a rarer 
group (hence more distant) and have broader lines (hence less accurate radial 
velocities).  Of particular interest among the  binary characteristics 
of intermediate/high mass stars are the frequency of
binary and multiple systems, the distribution of mass ratios, and the distribution of
separations (including the maximum separation).  Cepheids (5 M$_\odot$ stars) provide a 
number of tools for investigating binary properties, particularly using multiwavelength 
techniques. The next sections describe several studies which have the
over-arching goal of comparing their binary properties with those of low-mass stars.

\runinhead{HST Survey:}  We are conducting a survey of 69 of the brightest/nearest 
Cepheids with the Hubble Space Telescope (HST) Wide Field Camera 3 (WFC3), obtaining 
images in filters which transform to V and I.  The  goal is to look for companions as 
close as 100--200 AU.  The fields typically cover approximately 
0.1 pc = 20,000 AU, a standard expectation for the maximum dimension of binaries.  

\runinhead{HST Survey: $\eta$ Aql} has been known since an early IUE (International Ultraviolet Explorer Satellite) 
observation to have a hot companion. However no evidence of orbital motion has been found
in extensive radial velocity observations.  
Fig.~1 (left) shows the image of the Cepheid $\eta$ Aql, displaying a complex
point spread function (PSF). Comparison with other WFC3 images revealed the companion, resolved at about 7 o'clock.
Work is in progress on PSF correction. However for this relatively bright companion, we started 
with a simple subtraction.  When the image of the Cepheid T Mon (scaled and aligned) was subtracted
(Fig.~1 right),  the companion was  revealed clearly.  

\runinhead{HST Survey: Low-Mass Companions.}
The most difficult companions to identify are low-mass companions with a small mass ratio
relative to the
primary.  The second goal of the HST survey is to identify possible low-mass resolved companions.  
Fig.~2 (left) provides as an example, the V image of R Mus.  Possible companions  in the field
have been identified.  The V-(V-I) color magnitude diagram (Fig.~2 right) shows that only one star has 
the appropriate color--magnitude combination to be a likely companion.

%
\begin{figure}
\centering


\includegraphics[width=5.8cm]{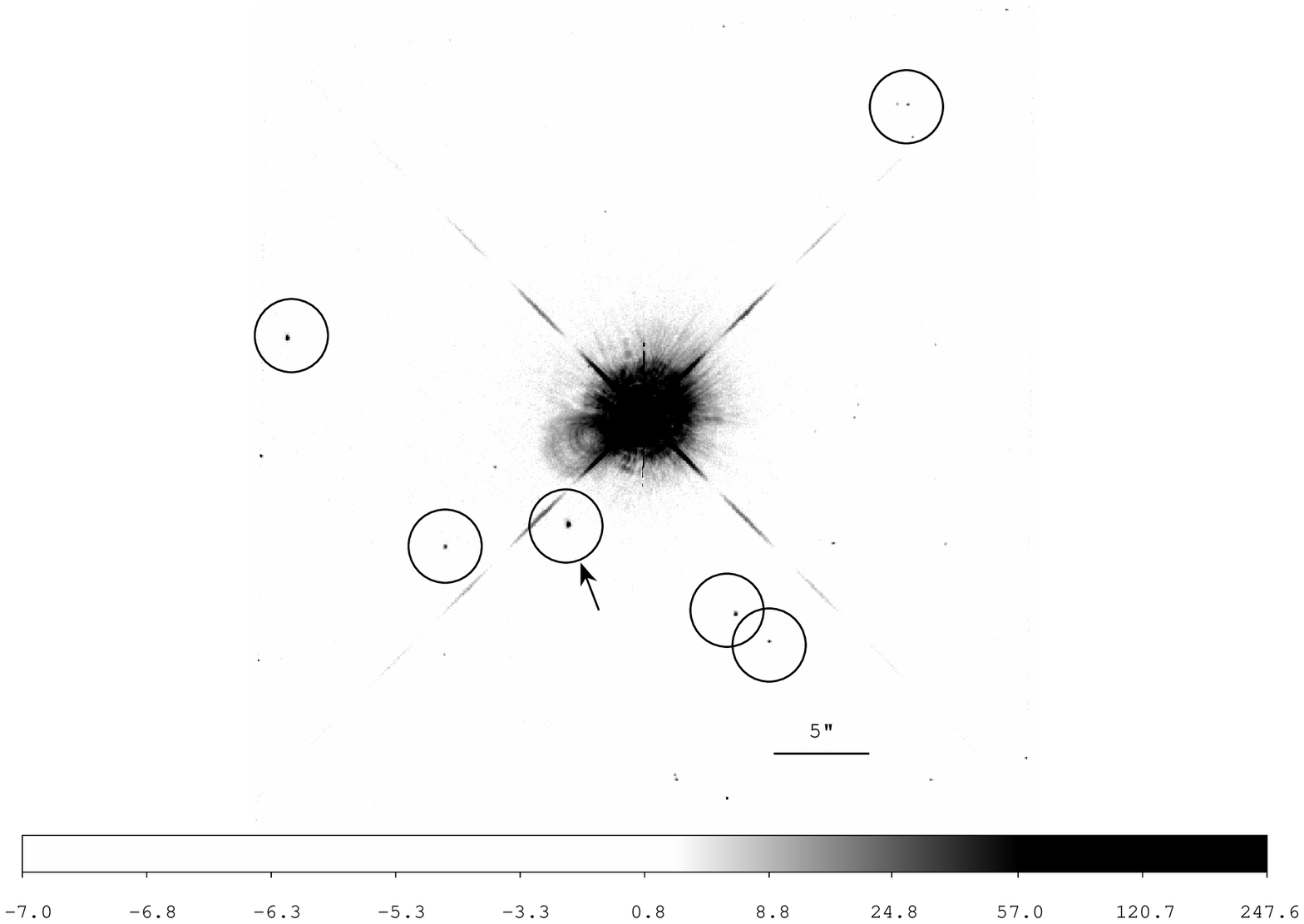}
\hspace{-0.05cm}
\includegraphics[width=5.8cm]{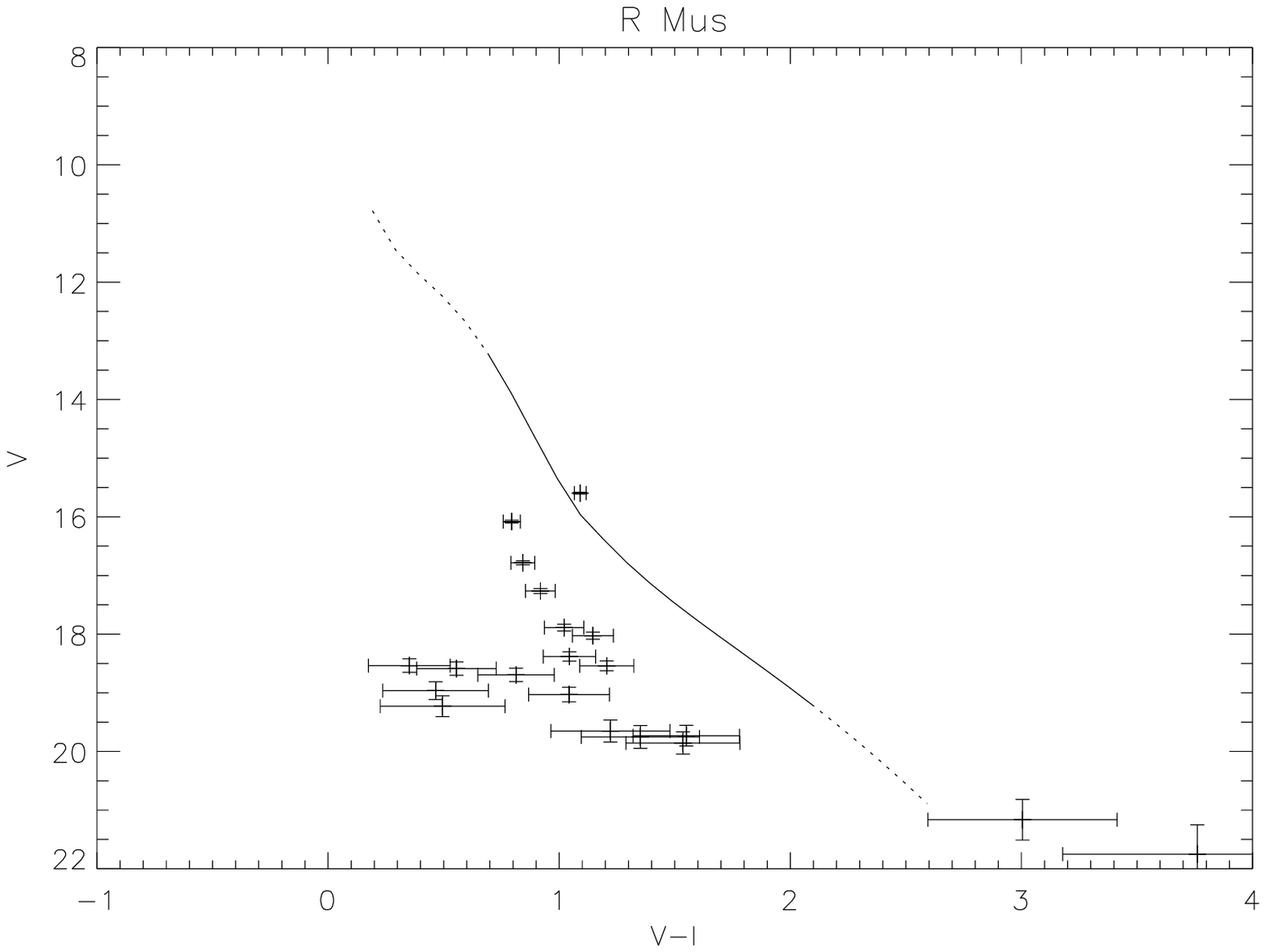}
%
%
\caption{Left: The V image of R Mus.  The brighter stars among the possible companions
have been circled, and the arrow indicates the probable companion from 
Fig.~2 (right). Right: The V-(V-I) diagram for faint stars in the R Mus field.  The
solid line is a ZAMS with the distance and reddening appropriate to the Cepheid. 
Dotted segments indicate stars too hot (hotter than mid-F spectral type) to produce X-rays or 
so cool that X-rays would be very difficult to detect.      }
\label{fig:2}       
\end{figure}



\runinhead{X-Rays from Low-Mass Companions.}  Another approach to determining the 
frequency of low-mass companions of high/intermediate mass stars uses X-ray observations.
Late B stars (as well as A stars) do not in general emit X-rays themselves, hence the 
X-rays that are occasionally found at the location of late B stars are thought 
to be produced by low-mass X-ray active companions (stars later than mid-F spectral type).  
We have developed a list of late B stars in the Carina Nebula cluster Tr 16 based on 
photometry and proper motions \cite{ev11a}.  The positions of these stars were 
compared with the locations of  X-ray source  in a Chandra ACIS image of the cluster, and the results
are shown in Fig.~3.  From the fraction of X-ray detected stars, we find that 39
percent of late B stars have low-mass companions with masses between 1.4 and 0.5 M$_\odot$.  
We note that a roughly equal fraction of {\it high} mass companions were detected in an 
IUE survey of Cepheids \cite{ev92}, making the combined fraction of 
5 M$_\odot$ stars with companions approximately three quarters.

\begin{figure}
\centering

\includegraphics[height=7cm]{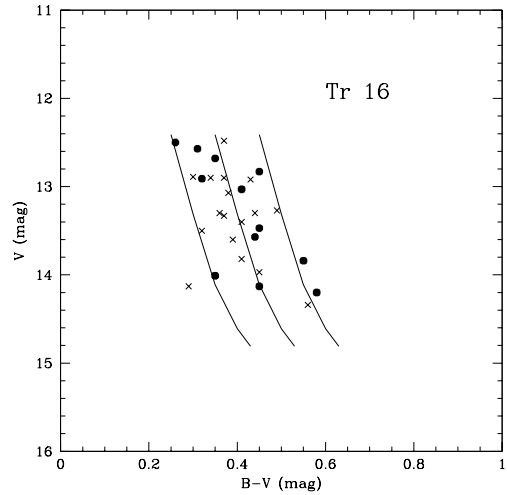}

\caption{The V--(B-V) colour-magnitude diagram of the late B stars in Tr 16 based on Cudworth proper 
motions.  Solid lines are the ZAMS with the distance and reddening of Tr~16 (center) 
with a range in E(B-V) of $\pm 0.1$ mag (left and right), which is the estimated 
dispersion in reddening in the cluster.  Dots are X-ray sources; crosses are
stars not detected in X-rays.  Reprinted from ApJS, 194, 13.}
\label{fig:3}       
\end{figure}

\section{Stellar Evolution: Masses}

 Observed mass determinations are based on spectroscopic orbits.  Considerable new velocity 
data for Cepheids are available from sources such as the Automated Spectrographic Telescope 
(AST; Joel Eaton) and the Moscow University group.  As an example, see the recent orbit of 
V350~Sgr \cite{ev11b}.   We also have a related ongoing project to use such data sources to 
search for low amplitude, long period orbits of Cepheids.

\begin{acknowledgement} Financial assistance for this work was provided by  Hubble grant 
GO-12215.01-A and the Chandra X-ray Center NASA contract NAS8-03060.
Fig.~1 was prepared with the help of Howard Bond and Fig.~2 with the help of Evan 
Tingle.

\end{acknowledgement}

\end{document}